\documentclass[12pt]{article}
 \usepackage{amsmath}
\usepackage{amssymb}
\usepackage{graphicx}
\begin{document}


\begin{center}
{\Large
	{\sc  On the Distribution of the Weighted Sum of Chi-Squared Variables}
}
\bigskip

 Ay\c{s}e \"Unsal \& Raymond Knopp 
\bigskip

{\it  
EURECOM, Sophia Antipolis, France \\
firstname.lastname@eurecom.fr
}
\end{center}
\bigskip

{\bf Abstract.} This paper presents the probability distribution function (p.d.f.) and cumulative distribution function (c.d.f.) of the weighted sum of central independent chi-squared random variables with non-zero weighs based on a method using moment generating functions. 
To obtain the p.d.f. and c.d.f. of such a function, we first derive the moment generating function of this weighted sum using the partial fractions decomposition or the residue method. The results cover the sum of two and three weighted chi-squared random variables, which can easily be adapted to more general cases. 


\bigskip\bigskip


\section{Introduction}
Statistical distributions come up in various application areas of the probability theory. Information theory, which studies and analyzes the fundamental limits of communication systems through probabilistic tools, is one of these areas where mathematical statistics plays an important role. For instance, in \textit{network information theory}, where the communication system is composed of multiple transmitters, multiple receivers or both, following normal distributions with various parameters, information theorists encounter a distribution in the form of a finite linear combination of independent central chi-square random variables in their analyses. 

In this work, we focus on the derivation of the p.d.f. and c.d.f. for such a linear function defined as follows
\begin{equation} \label{eq:def}
f(x)=\sum_{j} \lambda_j \mu_{(n_j)}^2, \; \mathrm{for} \; j=2,3,\cdots
\end{equation} where $\mu_{(n_j)}^2$'s denote independent chi-squared random variables with $n$ degrees of freedom and $\lambda_j$'s are non-zero and real factors. 
In this paper, we derive the \textit{exact} distribution of (\ref{eq:def}) for $j=2$ and $j=3$, through translating this expression into a sum of partial fractions. This method can easily be generalized to any value of $j$. 

\subsection{Related Work}

In \cite{r4}, the authors present the cumulative distribution function of a linear combination of independent central chi-square random variables through translation of its moment generating function into an infinite gamma series. The main difference between \cite{r4} and the current work is that in (\ref{eq:def}) the weighs represented by the terms $\lambda_j$'s are not limited to be positive. A less recent work on the same problem in \cite{r1}, considers the case where $\sum \lambda_j =1$ in addition to the condition of positive factors. \cite{r2} however, focuses on the distribution of the quadratic forms defined as $\sum_{i=1}^n \lambda_j (\mu_{j,i}+a_{j,i})^2$ where $a_{j,i} >0$. In \cite{r2}, the authors propose significance points for several values of $j$ and present two new approximations using the first three and four moments of the considered quadratic form. \cite{r5} and \cite{r6} are focused on series representations of quadratic forms for the vector consisting multivariate normal variables in different scenarios. In a quite related and rather recent study \cite{r7}, the authors characterize the distribution of positive quadratic forms of normal random variables deriving Laguerre expansions for the probability and cumulative distribution functions which is based on the inverse Laplace transforms.

\section{The Partial Fractions of the Moment Generating Function}
The random variables denoted $\mu_{(n_j)}^2$ in (\ref{eq:def}) has the following moment generating function of the chi-squared distribution with $n$ degrees of freedom
\begin{equation}
M(t)=(1-2t)^{-n/2}.
\end{equation} 
The sum of chi-squared variables that are weighted by the eigenvalues of the quadratic forms as given by (\ref{eq:def}) may occur in various scenarios one of which is the derivation of the mutual information function for finite $n$. When the weighs (i.e., $\lambda_j$'s) are equal to 1, sum of the chi-squared variables correspond to a gamma distribution with $n/2$ degrees of freedom. Here we have a special case of the gamma distribution since some of the factors are allowed to be negative and the gamma distribution is defined positive. Let us consider the simplest scenario for $j=2$, the corresponding moment generating function of the sum of two chi-squared distributed variables with weighs denoted by $\lambda_1$ and $\lambda_2$ is given as
\begin{equation} \label{mgf_gen}
M_g(t)=[1-2\lambda_1 t]^{-n/2} [1-2 \lambda_2t]^{-n/2} 
\end{equation} 
One could imagine $M_g(t)$ in a form of the sum of partial fractions as follows
\begin{equation}\label{mgf_Partial}
M_g(t)=\sum_{i=1}^{n}  \left(A_i [1-2\lambda_1 t]^{-n/2+i-1} +B_i[1-2\lambda_2 t]^{-n/2+i-1} \right)
\end{equation}
Multiplying both sides of (\ref{mgf_Partial}) by $[1-2\lambda_1 t]^{n/2}$, we have
\begin{equation} \label{mgf_partial_exp}
M_g(t)[1-2\lambda_1 t]^{n/2}= 
\sum_{i=1}^{n}  \left(A_i [1-2\lambda_1 t]^{i-1} +B_i[1-2\lambda_1 t]^{n/2}[1-2\lambda_2 t]^{-n/2+i-1} \right)
\end{equation}
Taking the derivatives of both sides of (\ref{mgf_partial_exp}) upto $i-1$, the partial fraction coefficients $A_i$ and $B_i$ are respectively derived as follows.
\begin{align} \label{eq:A}
A_i&=\frac{(-2\lambda_1)^{1-i}}{(i-1)!}\left.\frac{d^{i-1}}{dt^{i-1}}[1-2\lambda_1 t]^{n/2}M_g(t)\right|_{t=1/2\lambda_1} \nonumber \\
&=\frac{(\lambda_1/\lambda_2)^{1-i}}{(i-1)!}\left(\prod_{j=1}^{i-1} -\frac{n}{2}-(j-1)\right) (1-\lambda_2 /\lambda_1)^{-n/2-i+1}
\end{align}
\begin{align}\label{eq:B}
B_i&=\frac{(-2\lambda_2)^{1-i}}{(i-1)!}\left.\frac{d^{i-1}}{dt^{i-1}}[1-2\lambda_2 t]^{n/2}M_g(t)\right|_{t=1/2\lambda_2} \nonumber \\
&=\frac{(\lambda_2/\lambda_1)^{1-i}}{(i-1)!}\left(\prod_{j=1}^{i-1} -\frac{n}{2}-(j-1)\right)  (1-\lambda_1 /\lambda_2)^{-n/2-i+1}
\end{align}
The resulting probability distribution and cumulative distribution functions for ${j=2}$ and $x \geq 0$ are 
\begin{equation}\label{eq:pdf_j2}
f_2(x)=\sum_{i=1}^n \left(\frac{A_i}{\Gamma \left(\frac{n}{2}-i+1 \right) (2 \lambda_1)^{\frac{n}{2}-i+1}} \mathrm{e}^{-\frac{x}{2\lambda_1 i}}+\frac{B_i}{\Gamma \left(\frac{n}{2}-i+1 \right) (2 \lambda_2)^{\frac{n}{2}-i+1}} \mathrm{e}^{-\frac{x}{2\lambda_2 i}} \right)  x^{\frac{n}{2}-i},
\end{equation}
\begin{equation}\label{eq:cdf_j2}
F_2(x)=\sum_{i=1}^n \left(\frac{A_i}{\Gamma \left(\frac{n}{2}-i+1\right)} \gamma \left(\frac{n}{2}-i+1,\frac{x}{2 \lambda_1}\right)+\frac{B_i}{\Gamma \left(\frac{n}{2}-i+1\right)} \gamma \left(\frac{n}{2}-i+1,\frac{x}{2 \lambda_2}\right) \right),
\end{equation} 
respectively. $A_i$ and $B_i$ are respectively given by (\ref{eq:A}) and (\ref{eq:B}), where the gamma function $\Gamma(.)$ is defined as
\begin{equation}\label{eq:gamma}
\Gamma(a)=\int_{0}^{\infty}t^{a-1} \mathrm{e}^{-t}dt
\end{equation} and $\gamma(.)$ denoting the incomplete function that is
$\Gamma_x(a)=\int_{0}^{x}t^{a-1} \mathrm{e}^{-t}dt$.

\subsection{Three terms case}

Imagine the case where $j=3$ in (\ref{eq:def}). The corresponding moment generating function in this scenario denoted $M_{g_3}(t)$ becomes
\begin{equation}\label{eq:Mg3}
M_{g_3}(t)=[1-2\lambda_1 t]^{-n/2} [1-2 \lambda_2t]^{-n/2} [1-2 \lambda_3t]^{-n/2} 
\end{equation} (\ref{eq:Mg3}) can be rewritten through its partial fractions as follows
\begin{equation}
M_{g_3}(t)= \sum_{i=1}^{n/2} \left \{A_i [1-2\lambda_1 t]^{-n/2+i-1} +B_i [1-2\lambda_2 t]^{-n/2+i-1}+C_i [1-2\lambda_3 t]^{-n/2+i-1} \right\} 
\end{equation} where the terms $A_i$, $B_i$ and $C_i$ are respectively given by
\begin{align} \label{eq:Ai_3}
A_i&=\frac{(-2\lambda_1)^{1-i}}{(i-1)!}\left.\frac{d^{i-1}}{dt^{i-1}}[1-2\lambda_1 t]^{\frac{n}{2}}M_{g_3}(t)\right|_{t=1/2\lambda_1} \nonumber \\
&\overset{(a)}{=} \frac{(-2\lambda_1)^{1-i}}{(i-1)!}\sum_{k=0}^{i-1}\binom{i-1}{k}\frac{d^{i-1-k}}{dt^{i-1-k}}[1-2\lambda_2 t]^{i-1-k} \frac{d^{k}}{dt^{k}}[1-2\lambda_3 t]^{k}|_{t=1/2\lambda_1} \\
&=\frac{(\lambda_1/\lambda_2)^{1-i}}{(i-1)!}\sum_{k=0}^{i-1}\left[\binom{i-1}{k}\left(\prod_{j=1}^{i-1-k}-\frac{n}{2}-j+1\right) \left(1-\frac{\lambda_2}{\lambda_1}\right)^{-\frac{n}{2}-(i-1-k)}\right. \\
&\left.\;\;\;\;\;\;\;\;\;\;\;\;\;\;\;\;\;\;\;\;\;\;\;\;\;\;\;\;\;\; \left(\prod_{j=1}^{k}-\frac{n}{2}-j+1\right)\left(1-\frac{\lambda_3}{\lambda_1}\right)^{-\frac{n}{2}-k}\left(\frac{\lambda_3}{\lambda_2}\right)^k \right]
\end{align}
\begin{align}\label{eq:Bi_3}
B_i&=\frac{(-2\lambda_2)^{1-i}}{(i-1)!}\left.\frac{d^{i-1}}{dt^{i-1}}[1-2\lambda_2 t]^{n/2}M_{g_3}(t)\right|_{t=1/2\lambda_2} \nonumber \\
&\overset{(b)}{=}\frac{(-2\lambda_2)^{1-i}}{(i-1)!}\sum_{k=0}^{i-1}\binom{i-1}{k}\frac{d^{i-1-k}}{dt^{i-1-k}}[1-2\lambda_1 t]^{i-1-k} \frac{d^{k}}{dt^{k}}[1-2\lambda_3 t]^{k}|_{t=1/2\lambda_2}\\
&=\frac{(\lambda_2/\lambda_1)^{1-i}}{(i-1)!}\sum_{k=0}^{i-1}\left[\binom{i-1}{k}\left(\prod_{j=1}^{i-1-k}-\frac{n}{2}-j+1\right) \left(1-\frac{\lambda_1}{\lambda_2}\right)^{-\frac{n}{2}-(i-1-k)}\right. \\
&\left.\;\;\;\;\;\;\;\;\;\;\;\;\;\;\;\;\;\;\;\;\;\;\;\;\;\;\;\;\;\; \left(\prod_{j=1}^{k}-\frac{n}{2}-j+1\right)\left(1-\frac{\lambda_3}{\lambda_2}\right)^{-\frac{n}{2}-k}\left(\frac{\lambda_3}{\lambda_1}\right)^k \right]
\end{align}
\begin{align}\label{eq:Ci_3}
C_i&=\frac{(-2\lambda_3)^{1-i}}{(i-1)!}\left.\frac{d^{i-1}}{dt^{i-1}}[1-2\lambda_3 t]^{n/2}M_{g_3}(t)\right|_{t=1/2\lambda_3} \nonumber \\
&\overset{(c)}{=} \frac{(-2\lambda_3)^{1-i}}{(i-1)!}\sum_{k=0}^{i-1}\binom{i-1}{k}\frac{d^{i-1-k}}{dt^{i-1-k}}[1-2\lambda_1 t]^{i-1-k} \frac{d^{k}}{dt^{k}}[1-2\lambda_2 t]^{k}|_{t=1/2\lambda_3} \\
&=\frac{(\lambda_3/\lambda_1)^{1-i}}{(i-1)!}\sum_{k=0}^{i-1}\left[\binom{i-1}{k}\left(\prod_{j=1}^{i-1-k}-\frac{n}{2}-j+1\right) \left(1-\frac{\lambda_1}{\lambda_3}\right)^{-\frac{n}{2}-(i-1-k)}\right.\\
&\left.\;\;\;\;\;\;\;\;\;\;\;\;\;\;\;\;\;\;\;\;\;\;\;\;\;\;\;\;\;\; \left(\prod_{j=1}^{k}-\frac{n}{2}-j+1\right)\left(1-\frac{\lambda_2}{\lambda_3}\right)^{-\frac{n}{2}-k}\left(\frac{\lambda_2}{\lambda_1}\right)^k\right]
\end{align} In steps (a), (b) and (c), we used the following general Leibniz rule for the $n^{th}$ derivative of a product that is
\begin{equation}
(fg)^n(x)=\sum_{k=0}^{n} \binom{n}{k} f^{n-k}(x) g^{k}(x)
\end{equation}
Finally, the resulting probability distribution and cumulative distribution functions for ${j=3}$ are 
\begin{align}\label{eq:pdf_j3}
f_3(x)=\sum_{i=1}^n &\left(\frac{A_i}{\Gamma \left(\frac{n}{2}-i+1 \right) (2 \lambda_1)^{\frac{n}{2}-i+1}} \mathrm{e}^{-\frac{x}{2\lambda_1 i}}  x^{\frac{n}{2}-i} +\frac{B_i}{\Gamma \left(\frac{n}{2}-i+1 \right) (2 \lambda_2)^{\frac{n}{2}-i+1}} \mathrm{e}^{-\frac{x}{2\lambda_2 i}}  x^{\frac{n}{2}-i} \right. \nonumber \\
&\left.+\frac{C_i}{\Gamma \left(\frac{n}{2}-i+1 \right) (2 \lambda_3)^{\frac{n}{2}-i+1}} \mathrm{e}^{-\frac{x}{2\lambda_3 i}}   x^{\frac{n}{2}-i}\right),
\end{align}
\begin{align}\label{eq:cdf_j3}
F_3(x)=\sum_{i=1}^n & \left(\frac{A_i}{\Gamma \left(\frac{n}{2}-i+1\right)} \gamma \left(\frac{n}{2}-i+1,\frac{x}{2 \lambda_1}\right) +\frac{B_i}{\Gamma \left(\frac{n}{2}-i+1\right)} \gamma \left(\frac{n}{2}-i+1,\frac{x}{2 \lambda_2}\right) \right. \nonumber \\
&\left.+\frac{C_i}{\Gamma \left(\frac{n}{2}-i+1\right)} \gamma \left(\frac{n}{2}-i+1,\frac{x}{2 \lambda_3}\right) \right),
\end{align} 
for $x\geq 0$, respectively. The gamma $\Gamma(.)$ and incomplete gamma $\gamma(.)$ functions are reminded above. Partial fraction factors $A_i$, $B_i$ and $C_i$ are derived in (\ref{eq:Ai_3})-(\ref{eq:Ci_3}), respectively.

\section{Numerical Evaluation Results}

\begin{figure}[h]
\centering
\includegraphics[width=.5\linewidth]{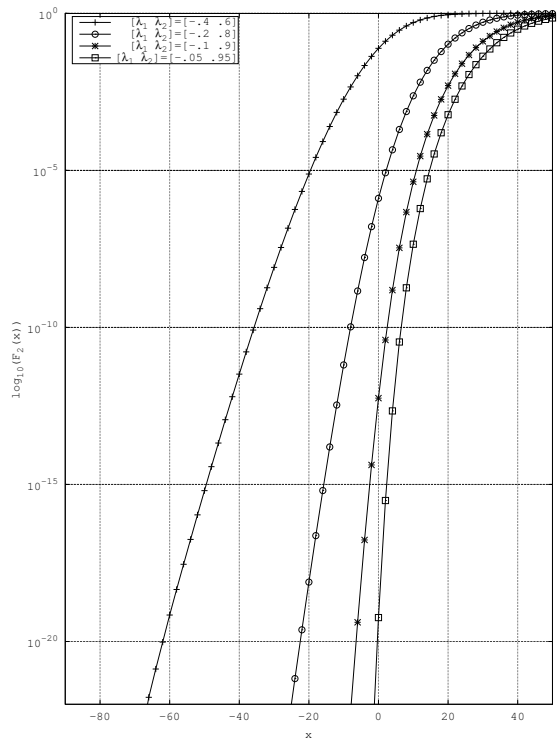}
\caption{Numerical evaluation of $F_2(x)$ for different pairs of $\lambda_1$, $\lambda_2$ and $n=50$.}
\label{fig:numeric2}
\end{figure}

Figure \ref{fig:numeric2} presents the c.d.f. given in (\ref{eq:cdf_j2}) with different positive and negative weighs as given in the legend. 




\begin{small}\end{small}


\begin{thebibliography}{9}
%
%

\bibitem{r1}
\textsc{Grad, A.} and \textsc{Solomon, H.} (1955). 
Distribution of Quadratic Forms and Some Applications
\textit{The Annals of Mathematical Statistics} \textbf{3} 464--477.

\bibitem{r2}
\textsc{Solomon, H.} and \textsc{Stephens, M. A.} (1977)
Distribution of a Sum of Weighted Chi-Square Variables \textit{Technical Report, Stanford University}


\bibitem{r4}
\textsc{Moschopoulos, P.G.} and \textsc{Canada, W.B.} (1984).
The Distribution Function of a Linear Combination of Chi-Squares \textit{Comp. \& Maths. with Appls.}
\textbf{10} 383--386.

\bibitem{r5}
\textsc{Kotz, S.} and \textsc{Johnson, N.L.} and \textsc{Boyd, D.W.}(1967). 
Series Representations of Distributions of Quadratic Forms in Normal Variables. I. Central Case
\textit{The Annals of Mathematical Statistics} Vol. 38, No. 3 823--837.

\bibitem{r6}
\textsc{Kotz, S.} and \textsc{Johnson, N.L.} and \textsc{Boyd, D.W.}(1967). 
Series Representations of Distributions of Quadratic Forms in Normal Variables. II. Non-Central Case
\textit{The Annals of Mathematical Statistics} Vol. 38, No. 3 838--848.

\bibitem{r7}
\textsc{Casta\~{n}o-Mart\'{i}nez, A.} and \textsc{L\'{o}pez-Bl\'{a}zquez, F.}(2005). 
Distribution of a sum of weighted noncentral chi-square variables
\textit{TEST} Vol. 14, No. 2 397--415.
\end{thebibliography}
\end{document}